\documentclass[final,twocolumn, 10pt]{IEEEtran}
\makeatletter

\let\proof\@undefined
\let\endproof\@undefined
\makeatother
\usepackage{epsfig}
\usepackage{amsbsy}
\usepackage{amsmath}
\usepackage{amsthm}
\usepackage{amssymb}
\usepackage{euscript}
\usepackage{citesort}
\usepackage{fancyhdr}
\usepackage{mathrsfs,bbm,dsfont}
\usepackage{amsfonts}
\usepackage[english]{babel}
\usepackage[latin1]{inputenc}
\usepackage{cite}
\input xy
\xyoption{all}
\usepackage{rotating}
\usepackage{url}

\newcounter{step}

\newcounter{stepit}

\newcommand{\mathsym}[1]{{}}

\newcommand{\modu}[1]{\left | #1  \right |}

\newcommand{\openone}{\mathbb{I}}
\renewcommand{\eqref}[1]{Eq.~(\ref{#1})}

\begin{document}
\title{Testing single-photon wave packets by~Hong-Ou-Mandel interference}
\author{Piotr Kolenderski and Konrad Banaszek
\thanks{\scriptsize
This work has been supported by the Polish budget funds for scientific research projects in years 2005-2008 and the European Commission under the Integrated Project Qubit Applications (QAP) funded by the IST directorate as Contract Number 015848.
\newline
\indent
Piotr Kolenderski and Konrad Banaszek are with
Institute of Physics, Nicolaus Copernicus University, Grudziadzka 5, 87-100 Toru{\'n}, Poland,
\newline
E-mail: $\tt kolenderski$@$\tt fizyka.umk.pl$
}}

\maketitle

\begin{abstract}
We discuss characterization of single-photon wave packets by measuring Hong-Ou-Mandel interference with a weak coherent pulse. A complete multimode calculation is presented and effects of multiphoton terms in the coherent field as well as the impact of source and detection imperfections are discussed.
\end{abstract}


\section{Introduction} Single photons are elementary building blocks in
optical implementations of quantum information processing, communication and cryptography \cite{Kok2007,Gisin2002,OBrien2003}. Apart from the simplest protocols that encode and process information using just one photon, more elaborate schemes require interactions between photons. Experimentally, the most accessible way to realize such interactions is to resort to multiphoton interference occurring in linear optical networks. Despite the limited range of transformations that can be achieved this way, supplementing linear optics with auxiliary sources and feed-forward operations yields universal quantum information processing capabilities \cite{Knill2001}.

Realization of multiphoton interactions by means of linear optics places stringent requirements on the modal characteristics of single photons. Interference effects require spatio-temporal indistinguishability of the single-photon wave packets at the input of the linear-optics networks. A sufficient test for this condition is the Hong-Ou-Mandel interference effect on a 50/50 beam splitter \cite{Hong1987}, when two incident indistinguishable photons are always found in the same output path. One possible realization of such a test is to use two photons originating from independent generation events \cite{Santori2002}. However, more insight into the modal structure of the generated photon can be gained by measuring two-photon interference with a suitably prepared weak coherent state. If such a state is picked off of a stronger coherent beam, its modal characteristics can be determined by measuring the remaining macroscopic part with one of many classical methods. This indirectly yields information about the single-photon wave packet, provided that the two-photon interference has been observed.

The purpose of this paper is to present a full multi-mode analysis of Hong-Ou-Mandel interference between a single photon and a coherent state. This allows us to address practical issues, such as the optimal amplitude of the employed coherent state: for its very small amplitude, the counting times would become prohibitively long; on the other hand, increasing the amplitude of the coherent state would enhance its two-photon and higher terms masking Hong-Ou-Mandel interference. Our calculation takes into account also other factors affecting the trade-off, such as imperfect preparation efficiency of single photons, non-unit detection efficiency and dark counts of the detectors. This allows us to link the two-photon interference visibility to the actual modal overlap between the single-photon wave packet and the coherent state, and determine the systematic error of such a measurement.


Nonclassical interference between a single photon and a weak coherent pulse has been observed by Rarity at al.~\cite{rarity1997}, who also presented a single-mode model. A measurement of two-photon interference with a suitably chosen collection of coherent pulse modes enabled the reconstruction of the complete single-photon density matrix in the spectral domain \cite{Wasilewski2007}. Here we extend previously used models to the fully multi-mode case and include the complete photon statistics of the coherent field.

\section{Beam splitter transformation}

We will consider here the spectral degree of freedom, corresponding to optical
fields confined to single-mode fibres. A generalization including spatial degrees of freedom is straightforward. We assume that the single-photon wave packet is prepared with a probability $p$, and is described by a density matrix $\hat\varrho(\omega,\omega')$. Thus the quantum state of the field entering the beam splitter reads:
\begin{equation}
    \hat{\varrho}_a=(1-p)|0\rangle\langle 0| +p \int d\omega
    \int d\omega'\varrho(\omega,\omega')
\hat{a}^\dagger(\omega)|0\rangle\langle0|\hat a(\omega')
\end{equation}
where $\hat{a}(\omega)$ and $\hat{a}^\dagger(\omega)$ are the annihilation and the
creation operators for a frequency $\omega$ at the corresponding input port of the
beam splitter. The coherent state is prepared in a mode characterized by a spectral amplitude $u(\omega)$, assumed to be normalized to one. The associated annihilation operator is therefore given by a superposition:
\begin{equation}
    \hat b=\int d\omega\ u^\ast(\omega) \hat b(\omega),
\end{equation}
where $\hat{b}(\omega)$ are annihilation operators of monochromatic modes entering the second input port of the beam splitter. We will use the following notation for a coherent state with an amplitude $\beta$ generated in the mode $u(\omega)$:
\begin{multline}
    |\beta u(\omega) \rangle_b = \hat D_b(\beta) |0\rangle= \\
    = \exp\left (-\frac{|\beta|^2}{2} + \int d\omega \, \beta u(\omega) \hat{b}^\dagger
    (\omega) \right) |0\rangle
\end{multline}
where  $\hat D_b(\beta)=\exp(\beta \hat b^\dagger -\beta^* \hat b ) $ is the
displacement operator.

The total input density matrix is given by the tensor product
$\hat{\varrho}_{\text{in}} = \hat{\varrho}_a \otimes |\beta u(\omega) \rangle_b
\langle \beta u(\omega) |$. This expression is transformed to the output density
matrix $\hat{\varrho}_{\text{out}}$ represented in terms of the outgoing modes $\hat
c(\omega)$ and $\hat d(\omega)$ leaving the beam splitter by substituting
monochromatic field operators according to:
\begin{equation}
  \hat a(\omega) = \frac{\hat c(\omega)-\hat d(\omega)}{\sqrt{2}} , \;\;\;
  \hat b(\omega) = \frac{\hat c(\omega)+\hat d(\omega)}{\sqrt{2}}
\end{equation}
We assume here a flat spectral characteristics of the beam splitter over the
relevant frequency range.

\section{Coincidence count rate}

The probability of a click on a binary detector monitoring
one of the outgoing beams is given by the expectation value of the following
normally ordered operator:
\begin{equation}
   \hat M_c =  \mathop{:} \hat{\openone} - \xi\exp\left(-\eta\int d\omega \,
    \hat{c}^\dagger(\omega)\hat{c}(\omega)\right):
\end{equation}
for the beam $c$ and analogously for the beam $d$. In the above formula, $1-\xi$ is the probability of a dark count, and $\eta$ is the quantum efficiency, assumed to be uniform across the spectrum of the detected fields. The coincidence count rate $R_C$ is given by the expectation value
\begin{equation}
    R_C = \text{Tr}(\hat{\varrho}_{\text{out}} : \hat{M}_c \otimes \hat{M}_d : ).
\end{equation}
This quantity can be conveniently expressed in terms of the expectation values of a two-parameter operator:
\begin{equation}
    \hat{Z}(\eta_c,\eta_d) =  :\exp\left(-\int d\omega[\eta_c
    \hat c^\dagger(\omega)\hat    c(\omega) + \eta_d\hat d^\dagger(\omega)\hat d(\omega)]
    \right):
\end{equation}
as
\begin{equation}\label{correlation}
    R_C=1-\xi[\langle \hat{Z}(\eta,0)\rangle +\langle \hat{Z}(0,\eta) \rangle]+
    \xi^2\langle \hat{Z}(\eta,\eta) \rangle.
\end{equation}
In order to evaluate the expectation value $\langle \hat{Z}(\eta_c,\eta_d)\rangle = \text{Tr}[\hat{\varrho}_{\text{out}} \hat{Z}(\eta_c,\eta_d)]$, it will be useful to write the output density matrix as:
\begin{multline}
        \hat{\varrho}_{\text{out}} = (1-p)\hat{\varrho}_{\text{coh}}+\\
        + \frac{p}{2}\int d\omega \int d\omega' \varrho(\omega,\omega')    [\hat{c}^\dagger(\omega) - \hat{d}^\dagger(\omega) ]    \hat{\varrho}_{\text{coh}}[\hat{c}(\omega') - \hat{d}(\omega') ]
\end{multline}
where
\begin{equation}
    \hat{\varrho}_{\text{coh}} = \left|\frac{\beta u(\omega)}{\sqrt{2}}    \right\rangle_c \left\langle
    \frac{\beta u(\omega)}{\sqrt{2}} \right| \otimes
    \left|\frac{\beta u(\omega)}{\sqrt{2}}
    \right\rangle_d \left\langle
    \frac{\beta u(\omega)}{\sqrt{2}} \right|
\end{equation}
describes a pair of coherent states in beams $c$ and $d$ prepared in a spectral mode described by the mode function $u(\omega)$, with equal amplitudes $\beta/\sqrt{2}$.

The expectation value $\langle \hat{Z}(\eta_c,\eta_d)\rangle$ involves then terms of the form $\text{Tr}[\hat{\varrho}_{\text{coh}}\hat{c}^\dagger(\omega) \hat{Z}(\eta_c,\eta_d)]$ and $\text{Tr}[\hat{\varrho}_{\text{coh}}\hat{c}^\dagger (\omega) \hat{Z}(\eta_c,\eta_d) \hat{c}(\omega')]$, where the operators for the beam $c$ and be also substituted by the field operators for the beam $d$. The operator products appearing in these expressions can be brought to the normally ordered form using the formula valid for any bosonic operator $\hat{c}$ \cite{Louisell1973}:
\begin{equation}
: \exp(-\eta \hat{c}^\dagger \hat{c}):\hat{c} = (1-\eta) \hat{c}
: \exp(-\eta \hat{c}^\dagger \hat{c}):
\end{equation}
Consequently:
\begin{equation}
    \hat c(\omega)\hat{Z}(\eta_c,\eta_d)  =  (1-\eta_c)\ \hat{Z}(\eta_c,\eta_d)\hat  c(\omega)
\end{equation}
and
\begin{multline}
    \hat c(\omega)\hat{Z}(\eta_c,\eta_d)\hat c^\dagger(\omega' )  = \\ =(1-\eta_c)\ \hat{Z}(\eta_c,\eta_d)\ + (1-\eta_c)^2\ \hat c^\dagger(\omega')\hat{Z}(\eta_c,\eta_d) \hat c(\omega)
\end{multline}
With the normally ordered expressions, the expectation value $\text{Tr}[\hat{\varrho}_{\text{out}} \hat{Z}(\eta_c,\eta_d)]$ can be easily calculated to be equal to:
\begin{multline}\label{zet:fin}
    Z(\eta_c,\eta_d)=\\
    \left(1-\frac{1}{2} (\eta _c+\eta _d) p+\frac{1}{4} (\eta _c-\eta _d)^2 p |\beta |^2 T \right) e^{- (\eta _c+\eta _d)|\alpha |^2 /2}
\end{multline}
where $T$ is the overlap of the single-photon wave packet with the mode function of the
probe coherent state:
\begin{equation}
    T=\int d\omega\ \int d\omega' u^\ast(\omega)\varrho(\omega,\omega')u(\omega').
\end{equation}

Substituting \eqref{zet:fin} into \eqref{correlation} we end up with the expression for the coincidence rate as a function of the overlap $T$ in the following form:
\begin{multline}\label{correlation:final}
    R_C(T) =1-\xi  \left(2 - \eta p + \frac{1}{2} \eta ^2  p  |\beta |^2 T \right)  e^{- \eta |\beta |^2/2}+\\
    +\xi^2 (1- \eta p) e^{-\eta  |\beta |^2}.
\end{multline}
The coincidence count rate measured as a function of the delay between the single-photon wave packet and the coherent pulse provides information on the spectral characteristics of the single-photon wave packet.

\section{Interference visibility}

\begin{figure*}[t]
  \epsfig{file=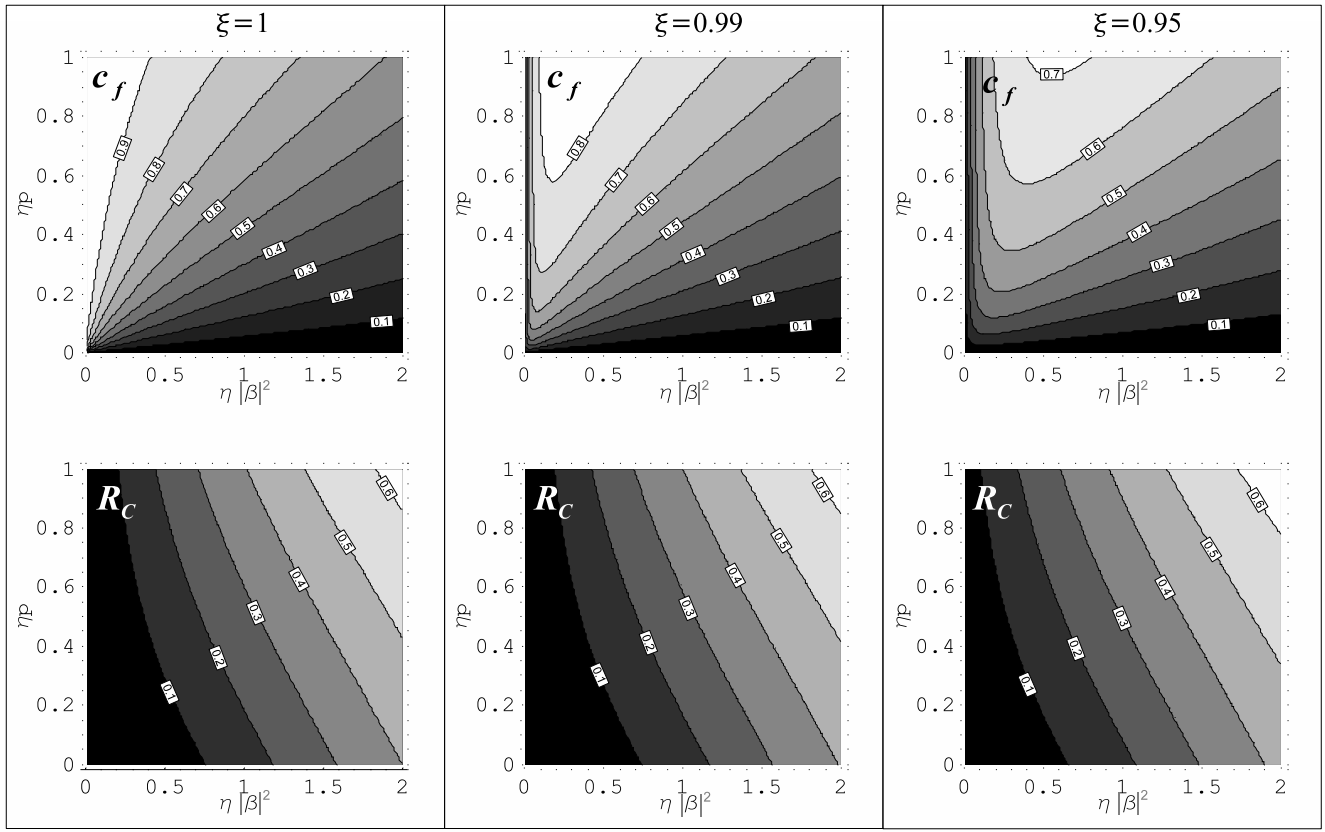,width=\linewidth}%
  \caption{Contour plots of the correction factor $c_f$ (top row) and the coincidence count rate $R_C$ (bottom row) as a function of the product of the generation probability and the detector efficiency  $\eta p$, and and measured coherent state intensity $\eta \modu{\beta}^2$.} \label{plot:cf}
\end{figure*}

The mode overlap $T$ can be recovered from \eqref{correlation:final}
by comparing the coincidence count rate with the case when the modes of the two incident beams are completely distinguishable, corresponding to $T=0$. This can be quantified with the help of the visibility of the Hong-Ou-Mandel dip, defined as:
\begin{equation}\label{visibility:def}
    V=\frac{R_C(0)-R_C(T)}{R_C(0)}
\end{equation}
After substitution of \eqref{correlation:final} into \eqref{visibility:def} we obtain
that the visibility is proportional to the mode
overlap $T$:
\begin{equation}
    V= c_f T
\end{equation}
where the proportionality constant $c_f$ is given by
\begin{equation}\label{cf}
    c_f=\frac{ \xi \eta^2 p  |\beta |^2e^{-\eta  |\beta |^2/2}}{2 [1- \xi (1-\eta p)
   e^{-\eta  |\beta |^2/2}] \left(1 - \xi e^{- \eta  |\beta |^2/2}\right)}
\end{equation}
and we will refer to it as the {\em correction factor}, dependent on the parameters of the setup. In
the idealized limit of $\eta=\xi=1$ and $\alpha \rightarrow 0$ the correction factor is equal to one.
In the first step, it is instructive to consider the case of no dark counts $\xi=1$ and a small coherent
amplitude, which allows one to apply an expansion in $\eta \modu{\beta}^2$. One then obtains:
\begin{equation}
    c_f \approx 1 - \left( \frac{1}{2\eta p} - \frac{1}{4} \right)\eta |\beta|^2
\end{equation}
which shows that the correction factor decreases with the increasing coherent
state effective amplitude. In the left panel of Fig.~\ref{plot:cf} we plot the correction factor $c_f$ and the coincidence count rate $R_C$ for $T=0$ as a function of $\eta|\beta|^2$ and $\eta p$ under the assumption of no dark counts $\xi=1$. Obviously, the coherent pulse amplitude needs to be non-zero to register any two-photon event at all, which results in a trade-off for its value.

The behavior of the correction factor $c_f$ is qualitatively different in the presence of dark counts characterized by $\xi < 1$, as seen in the center and right panels of Fig.~\ref{plot:cf}. This is because for very small coherent pulse amplitudes the interference dip becomes dominated by dark count events, and the visibility carries little information about the  mode overlap. Consequently, for a fixed product $\eta p$, there is a non-zero optimal coherent state amplitude that maximizes the correction factor $c_f$.
However, the correction factor never reaches the unit value in contrast to the regime free of dark counts. Fig.~\ref{plot:cf} shows that for detectors having 1\% and 5\% of dark counts the highest possible value of the correction factor is close to 0.8 and 0.7 respectively.

\begin{figure}[ht]
  \begin{center}
     \epsfig{file=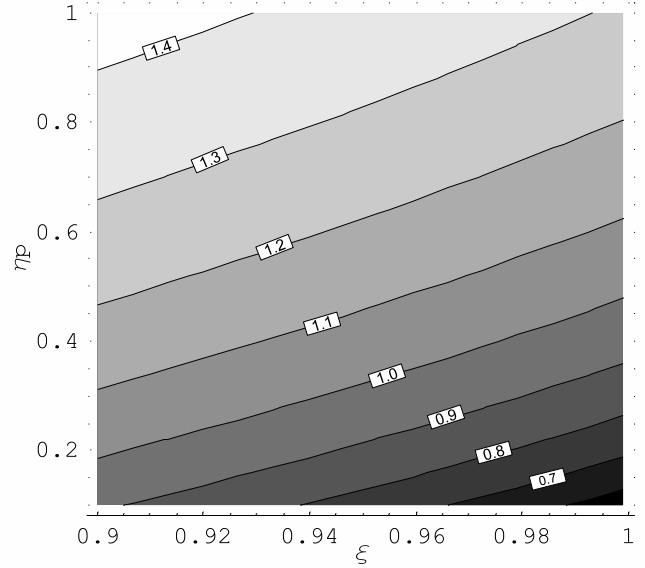,width=\linewidth}                
     \caption{The dependence of the effective coherent state intensity $\eta |\beta|^2$ on the dark count probability $\xi$ and the overall single photon production probability $\eta p$. }
    \label{plot:error}
  \end{center}
\end{figure}

The statistical uncertainty of the determined overlap $T$ is a function of both the correction factor, which attenuates the dependence of the measured visibility on $T$, and the count rate of coincidence events. In a simple
estimate, the relative uncertainty of $T$ can be written as:
\begin{equation}
\frac{\Delta T}{T} \propto \frac{1}{c_f \sqrt{R_C(0)}}
\end{equation}
The expression on the right hand side of the above formula can be maximized with respect
to $\eta|\beta|^2$ for fixed values of $\xi$ and $\eta p$. Results of this procedure are shown
in Fig.~\ref{plot:error}, which provides guidelines for selecting the optimal amplitude of the probe coherent pulse.
It is seen that the number of detected photons in the coherent pulse should be of the order of one.

\section{Conclusions}

The method of characterizing single-photon wave packets is very simple from the conceptual point of view. However, data obtained using an auxiliary coherent pulse to implement Hong-Ou-Mandel interference need to be corrected for multiphoton events. The optimal amplitude of the pulse has been found for a realistic range of parameters.

\bibliographystyle{IEEEtran}


\end{document}